\documentclass[aps,prl,preprint,showpacs]{revtex4}
\usepackage{amssymb}
\usepackage{bbm}
\usepackage{amsfonts}
\usepackage{amsmath}
\usepackage{mathrsfs}

\begin{document}
\title{Proper identification of the gluon spin}

\author{Xiang-Song Chen$^{1,2,3}$}
\email{cxs@hust.edu.cn}
\author{Wei-Min Sun$^{3,2}$}
\author{Fan Wang$^{3,2}$}
\author{T. Goldman$^4$}

\affiliation{$^1$Department of Physics, Huazhong
University of Science and Technology, Wuhan 430074, China\\
$^2$Kavli Institute for Theoretical Physics China, CAS, Beijing
100190, China \\
$^3$Department of Physics, Nanjing University, CPNPC, Nanjing
210093, China\\
$^4$Theoretical Division, Los Alamos National Laboratory, Los
Alamos, NM 87545, USA}

\date{\today}

\begin{abstract}
Properties of the recently proposed gauge-invariant gluon spin $S_g$
are studied and compared to the usually defined ``gluon
polarization'' $\Delta g$. By explicit 1-loop calculations in a
quark state, it is found that $S_g= \frac 59\Delta g$. Furthermore,
$\frac 45$ of $S_g$ can actually be identified as a ``static-field''
contribution and shown to cancel exactly an analogous static term in
the gluon orbital angular momentum $L_g$, leaving $S_g+L_g$
unaltered. These observations suggest that if properly identified,
the gluon contribution to the nucleon spin may be drastically
smaller than in the conventional wisdom.

\pacs{11.15.-q, 14.20.Dh, 12.38.-t}
\end{abstract}
\maketitle

In gauge theories, the basic physical notions such as gluon or
photon spin and orbital angular momentum suffer from a severe problem
with gauge invariance, because construction of these quantities
necessarily involves the canonical variable of the gauge field
$A_\mu$, which includes a gauge freedom. This problem has long
interfered with a coherent investigation of the nucleon spin structure,
and only recently, was a solution proposed in Refs.\cite{Chen08,Chen09},
which aroused considerable interest in the
community~\cite{Waka10,Cho10,Burk10,Lead11,Hatt11}. 
The key to the solution is the
extraction of a physical (gauge-invariant) part $\hat A_\mu$ out of
the gauge field $A_\mu$. (An analog of this idea actually has a long
history in gravity, in the attempt to extract the true gravitational
degrees of freedom out of the metric which also describes the
inertial effect~\cite{Chen11}). With this gauge-invariant field
variable $\hat A_\mu$, the gluon or photon spin {\em density} can be
easily defined gauge-invariantly as $\vec E\times \vec {\hat A} $,
which is the subject of close examination in this paper.

To start, we review and re-formulate the gauge-invariant
construction in Refs.\cite{Chen08,Chen09} for convenient use in the
present calculation. For an Abelian gauge field, the explicit
expression of $\hat A_\mu$ is \cite{Chen10d}:
\begin{equation}
\hat A_\mu(\vec x,t)= \frac{1}{\vec\partial ^2} \partial_i
F_{i\mu}(\vec x,t) =\int d^3 x' \frac {\partial'_i F_{i\mu}(\vec
x',t)} {-4\pi |\vec x-\vec x'|} \label{As}
\end{equation}
Here $F_{\mu\nu}=\partial_\mu A_\nu -\partial_\nu A_\mu$ is the
electromagnetic field strength. Greek indices run from 0 to 3, Latin
indices run from 1 to 3, and repeated indices are summed over (even
when they both appear raised or lowered). Gauge invariance of
$\hat A_\mu$ is evident from its expression in terms of $F_{\mu\nu}$,
and it can be easily verified that
\begin{subequations}
\label{hatA}
\begin{eqnarray}
\partial_\mu \hat A_\nu -\partial _\nu \hat A_\mu &=&\frac 1
{\vec \partial^2} (\partial_\mu\partial_i
F_{i\nu}-\partial_\nu\partial_i F_{i\mu})=F_{\mu\nu}\\
\partial_i \hat A_i &=&\frac 1
{\vec \partial^2} \partial_i \partial_k F_{ki}=0,
\end{eqnarray}
\end{subequations}
namely, that $\hat A_\mu$ has null spatial divergence, and produces the
full field strength $F^{\mu\nu}$. Thus,
\begin{equation}
\bar A_\mu \equiv A_\mu -\hat A_\mu=A_\mu-\frac{1}{\vec\partial ^2}
\partial_i F_{i\mu} \label{barAs}
\end{equation}
is the pure-gauge part of $A_\mu$, and is fully responsible for the
spatial divergence of $A_\mu$:
\begin{subequations}
\label{barA}
\begin{eqnarray}
\partial_\mu \bar A_\mu -\partial _\nu \bar A_\mu &=&0\\
\partial_i \bar A_i &=&\partial_i A_i
\end{eqnarray}
\end{subequations}

Equivalently, Eqs.({\ref{hatA}) and (\ref{barA}) can be taken as
the defining equations for $\hat A_\mu$ and $\bar A_\mu$, with
Eqs.(\ref{As}) and (\ref{barAs}) as their solutions, respectively. However,
one must be careful about boundary conditions. For the integration
in Eq.(\ref{As}) to be meaningful, the field strength $F_{\mu\nu}$
must vanish fast enough at infinity. It is natural to assign the
same boundary condition for the physical field $\hat A_\mu$; then
Eq.(\ref{As}) is the unique solution to Eqs.({\ref{hatA}), and Eq.
(\ref{barAs}) is the unique solution to Eqs.(\ref{barA}) under the
boundary condition that $\bar A_\mu$ approaches $A_\mu$ at infinity.
Note that $A_\mu$ or more precisely its pure-gauge part $\bar A_\mu$
may not vanish at infinity even if $F_{\mu\nu}$ does. However, if
they do (as typically obtains in perturbative calculations), the expressions
for $\hat A_\mu$ and $\bar A_\mu$ can be greatly simplified:
\begin{subequations}
\label{AA}
\begin{eqnarray}
\hat A_\mu &=&\frac{1}{\vec\partial ^2} \partial_i (\partial_i A_\mu
-\partial_\mu A_i)=A_\mu -\partial_\mu \frac{1}{\vec\partial ^2}
\partial_i A_i\\
\bar A_\mu &=&\partial_\mu \frac{1}{\vec\partial ^2}
\partial_i A_i
\end{eqnarray}
\end{subequations}

For the non-Abelian gluon field, $A_\mu \equiv A_\mu^a T^a$, with
$T^a$ the color matrix, the defining equations for $\hat A_\mu$ and
$\bar A_\mu$ are most elegantly arranged as \cite{Chen09,Chen10d}
\begin{subequations}
\label{nA}
\begin{eqnarray}
\bar F_{\mu\nu} &\equiv &\partial _\mu \bar A_\nu-\partial_\nu \bar
A_\mu +ig[\bar A_\mu,\bar A_\nu]=0 , \label{nA1}\\
\bar {\mathcal D}_i \hat A_i &\equiv& \partial_i \hat A_i +ig[\bar
A_i,\hat A_i]= 0 .\label{nA2}
\end{eqnarray}
\end{subequations}
That is, $\bar A_\mu$ is still a pure gauge, and the {\em gauge covariant}
spatial divergence of $\hat A_\mu$ vanishes. The physical field,
$\hat A_\mu $, (like the non-Abelian field strength $F_{\mu\nu}$) is
now gauge-covariant instead of gauge-invariant, and has to be solved
for perturbatively. The leading term is the same as in Eq. (\ref{As}).
At next-to-leading order, the expression is \cite{Chen10d}:
\begin{equation}
\hat A_\mu =\frac 1{\vec \partial^2} \partial_i F_{i\mu}+ig \frac
1{\vec
\partial^2} \{ [\frac 1{\vec \partial^2} \partial_k F_{ki},\partial_i\frac
1{\vec \partial^2} \partial_k F_{k\mu}-\partial_i A_\mu]
-\partial_i[A_i,\frac 1{\vec \partial^2} \partial_k F_{k\mu}]
+\partial_\mu [\frac 1{\vec \partial^2} \partial_k F_{ki} ,A_i]\} +
\mathcal {O}(g^2)
\end{equation}
This expression only requires that $F_{\mu\nu}$ vanish fast enough at
infinity. If $A_\mu$ does also, then the expressions for $\hat A_\mu$ and
$\bar A_\mu$ simplify to
\begin{subequations}
\label{gAA}
\begin{eqnarray}
\hat A_\mu &=&A_\mu -\bar A_\mu \\
\bar A_\mu &=&\partial_\mu \frac{1}{\vec\partial ^2}
\partial_i A_i +ig \{\partial_\mu \frac{1}{\vec\partial ^2}
[\partial_i \frac{1}{\vec\partial ^2}\partial_k A_k, A_i]
-\partial_i\frac{1}{\vec\partial ^2} [\partial_i
\frac{1}{\vec\partial ^2} \partial_k A_k ,\partial_\mu \frac
1{\vec\partial ^2} \partial_k A_k]\}+\mathcal {O}(g^2)
\end{eqnarray}
\end{subequations}

Separation of the physical and gauge degrees of freedom in $A_\mu$
can be of great use and convenience. A significant example is
the gauge-invariant, complete decomposition of the total QCD angular
momentum operator into four terms~\cite{Chen08,Chen09}:
\begin{eqnarray}
\vec J_{\rm QCD} &=& \int d^3 x \psi ^\dagger \frac 12 \vec \Sigma
\psi + \int d^3x \vec x \times \psi ^\dagger \frac 1i \vec {\bar D}
\psi +\int d^3x \vec E \times \vec {\hat A}+ \int d^3x \vec x\times
E_i \vec {\bar {\mathcal D}} \hat A_i \nonumber \\
&\equiv& \vec S_q +\vec L_q +\vec S_g +\vec L_g \label{QCD}.
\end{eqnarray}

Here $\bar D_\mu \equiv \partial_\mu +ig \bar A_\mu$ is the
pure-gauge covariant derivative for the quark field, and $\bar
{\mathcal D}_\mu \equiv \partial_\mu +ig [\bar A_\mu,~]$ is the
pure-gauge covariant derivative for a field in the adjoint
representation. For an Abelian theory, $\hat A_\mu$ is gauge invariant
and does not need a covariant derivative, so the decomposition of
the total QED angular momentum operator is a little simpler (in the
sector of photon orbital angular momentum):
\begin{eqnarray}
\vec J_{\rm QED} &=& \int d^3 x \psi ^\dagger \frac 12 \vec \Sigma
\psi + \int d^3x \vec x \times \psi ^\dagger \frac 1i \vec {\bar D}
\psi +\int d^3x \vec E \times \vec {\hat A}+ \int d^3x \vec x\times
E_i \vec \partial \hat A_i \nonumber \\
&\equiv& \vec S_e +\vec L_e +\vec S_\gamma +\vec L_\gamma
\label{QED}.
\end{eqnarray}
[A remarkable (and somewhat mysterious) fact is that using the ordinary
derivative can also lead to a gauge-invariant gluon orbital angular
momentum, which however dictates a very specific definition of $\hat
A_\mu$. See Ref. \cite{Chen08} for a detailed discussion.]

Note that Eqs.(\ref{QCD}) and (\ref{QED}) provide {\em operational}
decompositions: Given the explicit expressions of $\hat A_\mu$ and
$\bar A_\mu$, we can straightforwardly make calculations with the
operators in Eqs.(\ref{QCD}) and (\ref{QED}). We will concentrate here
on the gauge-invariant ``gluon spin'', $\vec S_g$. In comparison, the
widely discussed ``gluon polarization'' $\Delta g$ is related to the
gauge-dependent operator $\int d^3x \vec E \times \vec A$ in the
light-cone gauge~\cite{Jaff96}. The aim of this paper is to investigate
the properties of $\vec S_g$, and to make a quantitative comparison
with $\Delta g$. To make the conclusion as concrete as possible, we
take the simplest non-trivial example of a 1-loop calculation in an
on-shell quark state. As we will show, $S_g$ is typically much
smaller than $\Delta g$. Moreover, a remarkable and physically
appealing feature of our gauge-invariant definitions in Eq.(\ref{QCD})
is that the ``static field'' does not contribute to the
total gluon angular angular momentum, $S_g+L_g$. [The same feature
holds for the total photon angular momentum, $S_\gamma+ L_\gamma$).
If the static pieces are subtracted altogether from $S_g$ and $L_g$
(which leaves the sum, $S_g+L_g$, unaltered), the remaining ``dynamic''
gluon spin is only $\frac 19$ of $\Delta g$.

Since $\vec S_g$ and $\vec L_g$ are explicitly gauge-invariant
operators, the calculation can be performed in any gauge for
convenience. The covariant gauge has the simplest Feynman rules. But
for our purpose the Coulomb gauge $\vec \partial \cdot \vec A=0$ is
the most convenient: Eqs.(\ref{gAA}) indicate that as $\vec
\partial \cdot \vec A$, we have $\bar A_\mu =0$ and $\hat A_\mu
=A_\mu$ in this gauge.

For the 1-loop calculation of the gluon matrix element in a quark state,
the gluon field behaves like eight independent Abelian fields.
Consider a quark state, $|p\sigma\rangle$, with momentum $p$ and
polarization $\sigma$ along the third axis. At 1-loop order one
finds~\cite{Hood99}
\begin{equation}
\Delta g \equiv \langle p\sigma|\int d^3x (\vec E \times \vec A)_3
|p\sigma \rangle _{A^+ =0}=\sigma \cdot 2\frac{\alpha_s}{\pi}
\ln{\frac {Q^2}{m^2}}
\end{equation}
where $Q^2$ and $m^2$ are the untraviolet and infrared cutoffs,
respectively. For comparison, our gauge-invariant ``gluon spin''
leads to
\begin{eqnarray}
S_g&\equiv& \langle p \sigma |\int d^3x (\vec E \times \vec {\hat
A})_3
|p \sigma \rangle \nonumber\\
&=&\langle p\sigma|\int d^3x (\vec E \times \vec A)_3
|p\sigma\rangle _{\vec \partial \cdot \vec A =0}=\frac 59\Delta g.
\end{eqnarray}

Here we have used the important and convenient relation that
the matrix element of the {\em gauge-invariant} operator $\vec
E\times \vec {\hat A}$ (in any gauge) is the same as that of the
{\em gauge-dependent} operator $\vec E\times \vec A$ in the Coulomb
gauge. (The reader should keep in mind that $\Delta g$ has been
defined only in light-cone gauge, whereas we have only {\em chosen}
to evaluate our {\em gauge-invariant} operator in Coulomb gauge.)

We see that $S_g$ is much smaller than $\Delta g$. The reason
can be traced to the fact that $S_g$ is constructed solely with the
``physical'' gluon field, while $\Delta g$ is calculated in the
light-cone gauge in which $A_\mu$ contains both physical and
pure-gauge components; thus $\Delta g$ includes a non-physical
pure-gauge contribution. This suggests that $S_g$ is a more physical
and reasonable definition of the gluon spin than $\Delta g$. And
indeed, $\Delta g$ leads to a spuriously large gluon content in a
parent quark state. A rather heuristic way to see this is by
renormalizing the divergent $S_g$ and $\Delta g$ in a very specific
way, namely, by choosing the ultraviolet cutoff $Q^2$ to be the same
as the scale for the running coupling constant:
\begin{equation}
\alpha_s(Q^2) =\frac{g^2(Q^2)}{4\pi}=\frac{12 \pi}{(33-2n_f)\ln
(Q^2/\Lambda^2)}.
\end{equation}
If the quark flavor $n_f$ is set to $3$, then at large $Q^2$ we find
$\Delta g=\frac 89 \sigma $, while $S_g\simeq 0.5\sigma$.

We now derive the major conclusion of this paper, namely that the
gauge-invariant gluon or photon spin $\vec E\times \vec {\hat A}$
can be further separated into two gauge-invariant terms, one of which,
a ``static-field'' term, exactly cancels an analogous term in the gluon
or photon orbital angular momentum. The point is that the separation
$A_\mu=\hat A_\mu+\bar A_\mu$ allows one to split the electric field
$\vec E$ into {\em gauge-invariant} (in Abelian case) or {\em
gauge-covariant} (in non-Abelian case) pieces. (Such splitting is
not possible with the full $A_\mu$.)

For the simpler Abelian case, we have:
\begin{equation}
\vec E=-\partial_t \vec A -\vec \partial A^0 =-\partial_t \vec {\hat
A} -\vec \partial \hat A^0 \equiv \vec E^{\rm dy}+\vec E^{\rm st}.
\label{E+E}
\end{equation}
We call $\vec E^{\rm dy}\equiv -\partial_t \vec {\hat A}$ the
``dynamic'' field and $\vec E^{\rm st}\equiv -\vec \partial \hat
A^0$ the ``static'' field. [We apologize that the word ``static'' is
not very pertinent, because $-\vec \partial \hat A^0$ can as well be
time-dependent. By ``static'' we mean exactly ``can-be-static'',
i.e., $-\vec \partial \hat A^0$ can survive for a static
configuration, while the ``dynamic'' field $-\partial_t \vec {\hat
A}$ cannot. This notation will prove quite illuminating for the
non-Abelian field.] As promised, $\vec E^{\rm dy}$ and $\vec E^{\rm
st}$ are separately gauge-invariant, since they are constructed
with the gauge-invariant physical fields $\vec {\hat A}$ and
$\hat A^0$, respectively. Now the ``static'' and
``dynamic'' terms of the photon spin $\vec S_\gamma$ can be defined
as
\begin{subequations}
\begin{eqnarray}
\vec S^{\rm st}_\gamma\equiv\int d^3x \vec E^{\rm st} \times \vec
{\hat A}=\int d^3x (-\vec \partial \hat A^0) \times \vec {\hat A}\\
\vec S^{\rm dy}_\gamma\equiv\int d^3x \vec E^{\rm dy} \times \vec
{\hat A}=\int d^3x (-\partial_t\vec {\hat A}) \times \vec {\hat A}
\end{eqnarray}
\end{subequations}

Similarly, we can define ``static'' and ``dynamic'' terms of the
photon orbital angular momentum:
\begin{subequations}
\begin{eqnarray}
\vec L^{\rm st}_\gamma\equiv\int d^3x \vec x \times E^{\rm st}_i
\vec \partial {\hat A}_i=\int d^3x \vec x \times (-\partial_i \hat
A^0) \vec \partial {\hat A}_i\\
\vec L^{\rm dy}_\gamma\equiv\int d^3x \vec x \times E^{\rm dy}_i
\vec \partial {\hat A}_i=\int d^3x \vec x \times (-\partial_t \hat
A_i) \vec \partial {\hat A}_i
\end{eqnarray}
\end{subequations}

The ``static'' terms $\vec S^{\rm st}_\gamma$ and $\vec L^{\rm
st}_\gamma$ are not zero individually, but a little algebra shows
that the sum $\vec S^{\rm st}_\gamma+\vec L^{\rm st}_\gamma$
always vanishes:
\begin{subequations}
\label{st}
\begin{eqnarray}
\vec S^{\rm st}_\gamma&=&\int d^3x (-\vec \partial \hat A^0) \times
\vec
{\hat A}=\int d^3x \hat A^0 \vec \partial \times \vec {\hat A} \\
\vec L^{\rm st}_\gamma &=&\int d^3x (-\partial_i \hat A^0) \vec x
\times \vec \partial \hat  A_i = \int d^3x \hat A^0 (\partial_i \vec
x) \times \vec\partial \hat A_i + \int d^3x\vec x \times
\vec \partial (\partial_i \hat A_i) \nonumber \\
&=& -\int d^3x \hat A^0\vec\partial\times \vec {\hat A} =-\vec
S^{\rm st}_\gamma
\end{eqnarray}
\end{subequations}

Thus, it is not very meaningful to count $\vec S^{\rm st}_\gamma$ as
part of the photon spin and $\vec L^{\rm st}_\gamma=-\vec S^{\rm
st}_\gamma$ as part of the photon orbital angular momentum, and we
can wisely subtract these two canceling terms altogether from $\vec
S_\gamma$ and $\vec L_\gamma$. (This cancelation is also discussed by
Wakamatsu \cite{Waka10}.)

It is worthwhile to remark that the vanishing of ``static'' angular
momentum is masked if one defines the total photon angular momentum
as $\int d^3 x \vec x \times (\vec E\times \vec B)$. Similarly, by
defining photon momentum as $\int d^3 x \vec E\times \vec B$, ones
does not see the vanishing of a ``static'' term either, which
however shows up clearly if one defines the photon momentum as $\vec
P_\gamma \equiv \int d^3x E_i \vec \partial \hat A_i$:
\begin{eqnarray}
\vec P^{\rm st}_\gamma=\int d^3x E^{\rm st}_i \vec \partial \hat
A_i= \int d^3x (-\partial_i \hat A^0) \vec \partial \hat A_i =\int
d^3x \hat A^0 \vec \partial (\partial_i \hat A_i)=0.
\end{eqnarray}

Separation of the gluon spin $\vec S_g$, or essentially the
non-Abelian color electric field $\vec E$, is much more involved
due to non-linearity, but a satisfactory separation does exist:
\begin{eqnarray}
\vec E&=&-\partial_t\vec A-\vec \nabla A^0+ig [\vec A, A^0]
\nonumber \\
&=&-\bar {\mathcal D}_t \vec {\hat A}-\vec {\bar {\mathcal D}} \hat
A^0+ig [\vec {\hat A},
\hat A^0]\nonumber \\
&\equiv& \vec E^{\rm dy}+\vec E^{\rm st}+\vec E^{\rm nl} \label{YM}
\end{eqnarray}
In arranging the first line into the second line, we have used the
pure-gauge condition $ \vec {\bar E} \equiv-\partial_t\vec {\bar
A}-\vec \partial \bar A^0+ig [\vec {\bar A}, \bar A^0]=0$. In
addition to the ``dynamic'' term $\vec E^{\rm dy}\equiv -\bar
{\mathcal D}_t \vec {\hat A}$ and the ``static'' term $\vec E^{\rm
st}\equiv -\vec {\bar {\mathcal D}} \hat A^0$, we find now a
non-linear term $\vec E^{\rm nl}\equiv  ig [\vec {\hat A}, \hat
A^0]$. All these three terms are {\em individually} gauge-covariant.
In consequence, we can separate the gluon spin $\vec S_g$ into three
gauge-invariant parts:
\begin{subequations}
\begin{eqnarray}
\vec S^{\rm st}_g\equiv\int d^3x \vec E^{\rm st} \times \vec {\hat
A}=\int d^3x (-\vec {\bar {\mathcal D}} \hat A^0) \times \vec
{\hat A}\\
\vec S^{\rm dy}_g\equiv\int d^3x \vec E^{\rm dy} \times \vec
{\hat A}=\int d^3x ( -\bar {\mathcal D}_t \vec {\hat A}) \times \vec {\hat A}\\
\vec S^{\rm nl}_g\equiv\int d^3x \vec E^{\rm nl} \times \vec {\hat
A}=\int d^3x ig [\vec {\hat A}, \hat A^0] \times \vec {\hat A}
\end{eqnarray}
\end{subequations}

A similar gauge-invariant separation applies to the gluon orbital
angular momentum:
\begin{subequations}
\begin{eqnarray}
\vec L^{\rm st}_g\equiv\int d^3x \vec x \times E^{\rm st}_i \vec
{\bar {\mathcal D}} {\hat A}_i=\int d^3x \vec x \times (-\bar
{\mathcal D}_i \hat
A^0) \vec {\bar {\mathcal D}} {\hat A}_i\\
\vec L^{\rm dy}_g\equiv\int d^3x \vec x \times E^{\rm dy}_i \vec
{\bar {\mathcal D}} {\hat A}_i=\int d^3x \vec x \times (-\bar
{\mathcal D}_t \hat
A_i) \vec {\bar {\mathcal D}} {\hat A}_i\\
\vec L^{\rm nl}_g\equiv\int d^3x \vec x \times E^{\rm nl}_i \vec
{\bar {\mathcal D}} {\hat A}_i=\int d^3x \vec x \times ig[\hat A_i,
\hat A_0]\vec {\bar {\mathcal D}} {\hat A}_i
\end{eqnarray}
\end{subequations}

Although the expressions are much more complicated, we can still
prove that the total ``static'' angular momentum $\vec S^{\rm st}_g+
\vec L^{\rm st}_g$ vanishes identically. The easiest way to see this
is to work in the Coulomb gauge, which gives $\bar A_\mu=0$ and
$\hat A_\mu =A_\mu$, thus $\vec S^{\rm st}_g$ and $\vec L^{\rm st}_g$
reduce to the same expressions as for the photon. Then, the same
procedure as in Eqs. (\ref{st})  shows $(\vec S^{\rm st}_g+\vec
L^{\rm st}_g)_{\vec \partial \cdot \vec A=0}=0$. But since $(\vec
S^{\rm st}_g+\vec L^{\rm st}_g)$ is gauge invariant, it is
identically zero in any gauge. The same reasoning shows that the
``static'' gluon momentum $\vec P_g^{\rm st} \equiv \int d^3x E^{\rm
st}_i \vec {\bar {\mathcal D}} \hat A_i\equiv 0$. But again, the
vanishing of ``static'' gluon momentum and angular momentum are
masked if they are defined through the Poynting vector $\vec E\times
\vec B$.

Discarding the not-so-meaningful static term, we resume the
1-loop calculation and see how much ``essential gluon spin'' is
left in a parent quark state. At 1-loop order the non-linear term
$\vec S^{\rm nl}_g$ does not contribute. The remaining ``dynamic
gluon spin'' $\vec S^{\rm dy}_g$ is found to contribute
\begin{eqnarray}
S^{\rm dy}_g&\equiv& \langle p \sigma |\int d^3x (\vec E^{\rm dy}
\times \vec {\hat A})_3
|p \sigma \rangle \nonumber\\
&=&\langle p\sigma|\int d^3x (\vec E^{\rm dy} \times \vec A)_3
|p\sigma\rangle _{\vec \partial \cdot \vec A =0} \nonumber \\
&=&\frac 15 S_g=\frac 19 \Delta g=\sigma \cdot \frac
29\frac{\alpha_s}{\pi} \ln{\frac {Q^2}{m^2}}.
\end{eqnarray}

Quite remarkably, we see that $S^{\rm dy}_g$ is largely negligible
compared to $S_g$ or $\Delta g$, e.g., the specific renormalization
found by choosing the ultraviolet cutoff $Q^2$ to be the same as the
scale for $\alpha_s(Q^2)$ gives $S^{\rm dy}_g\simeq 0.1\sigma$.
To put this another way: the sizable ``gluon spin'', $S_g$, is mostly
a ``static-field'' contribution which is canceled exactly by the same
``static-field'' term in the ``gluon orbital angular momentum'', $L_g$.
{\em Significantly, this suggests that, defined properly, the gluon
contribution to the nucleon spin may be drastically smaller than in
the conventional wisdom.} In Ref.\cite{Chen09}, a similar feature has
been described for the gluon contribution to the nucleon momentum.

Although the Abelian and non-Abelian gauge fields share the same
property that a ``static'' term can be shown to vanish for the total
momentum and angular momentum, they do display a crucial difference
as well: For the non-Abelian case, the non-linear field $\vec E^{\rm nl}$
can survive for a time-independent configuration, which therefore can
possess non-trivial momentum and angular momentum. In this regard,
it is useful to separate the non-Abelian $\vec E$ into just two
terms: $\vec E= \vec E^{\rm st}+\vec E^{\rm nldy}$, where $\vec
E^{\rm nldy}=\vec E^{\rm nl}+\vec E^{\rm dy}$ is the
``non-linear-dynamic'' field, and can be conveniently written as
\begin{equation}
\vec E^{\rm nldy}=-\bar {\mathcal D}_t \vec {\hat A}+ig [\vec {\hat
A}, \hat A^0] =-\mathcal D_t \vec {\hat A}.
\end{equation}
Here $D_t=\partial_t +ig [A^0, ~]$ is the complete (versus
pure-gauge) covariant derivative. Accordingly, we can define the
``non-linear-dynamic'' gluon spin and orbital angular momentum as
\begin{subequations}
\begin{eqnarray}
\vec S^{\rm nldy}_g &\equiv& \int d^3x \vec E^{\rm nldy} \times \vec
{\hat A}=\int d^3x ( -\mathcal D_t \vec {\hat A}) \times \vec {\hat
A}\\
\vec L^{\rm nldy}_g&\equiv&\int d^3x \vec x \times E^{\rm nldy}_i
\vec {\bar {\mathcal D}} {\hat A}_i =\int d^3x \vec x \times
(-\mathcal D_t \hat A_i) \vec {\bar {\mathcal D}} {\hat A}_i
\end{eqnarray}
\end{subequations}
These are the non-trivial quantities which do not cancel against
each other. We can refine the decomposition in Eq. (\ref{QCD})
to be
\begin{equation}
\vec J_{\rm QCD} = \vec S_q +\vec L_q +
\vec S^{\rm nldy}_g +\vec L^{\rm nldy}_g .
\end{equation}

In closing, we comment on another possibly important application of
separating $\vec E$ into gauge-invariant/covariant pieces, regarding
the construction of novel parton distribution functions. In
Ref. \cite{Chen09}, we have remarked how the separation $A_\mu\equiv
\hat A_\mu+\bar A_\mu$ provides for a new definition of the polarized
gluon parton distribution function (PDF)
which corresponds to the gauge-invariant gluon spin, $\vec S_g$.
Analogously, the further separation, $\vec E= \vec E^{\rm st}+\vec
E^{\rm nldy}$, facilitates further the definition of a new PDF which
corresponds to the ``non-linear-dynamic gluon spin,'' $\vec S^{\rm
nldy}_g$. One may perform the same study for the gluon momentum,
but it is less significant because only the integrands are different in
$\vec P_g \equiv \int d^3x E_i \vec {\bar {\mathcal D}} \hat A_i= \int
d^3x E_i ^{\rm nldy}\vec {\bar {\mathcal D}} \hat A_i \equiv \vec P_g^{\rm nldy}$.

This work is supported by the China NSF under Grants No. 10875082
and No. 11035003, and by the U.S. DOE under Contract No.
DE-AC52-06NA25396. XSC is also supported by the NCET Program of the
China Education Department.

\end{document}